\pdfoutput=1
\documentclass[twocolumn,epjc3]{svjour3}
\usepackage{graphicx}
\usepackage{xcolor}
\usepackage{SIunits}
\usepackage{multirow}
\usepackage{setspace}
\usepackage[percent]{overpic}
\usepackage{contour}
\usepackage[T1]{fontenc} 

\usepackage{lineno}
\usepackage[colorlinks,citecolor=blue,urlcolor=blue]{hyperref}
\journalname{Eur. Phys. J. C}

\begin{document}
\title{Development of very-thick transparent GEMs with wavelength-shifting capability for noble element TPCs}

\author{M.~Ku\'zniak\thanksref{addr1,e1}
        \and D.~Gonz\'alez-D\'iaz\thanksref{addr2,e2}
        \and P.~Amedo\thanksref{addr2}
        \and C.\,D.\,R.~Azevedo\thanksref{addr3}
        \and D.\,J.~Fern\'andez-Posada\thanksref{addr2}
        \and M.~Ku\'zwa\thanksref{addr1}
        \and S.~Leardini\thanksref{addr2}
        \and A.~Leonhardt\thanksref{addr8}
        \and T.~\L ęcki\thanksref{addr4}
        \and L.~Manzanillas\thanksref{addr5}
        \and D.~Muenstermann\thanksref{addr6}
        \and G.~Nieradka\thanksref{addr1}
        \and R.~de~Oliveira\thanksref{addr7}
        \and T.\,R.~Pollmann\thanksref{addr8,currentNikhef}
        \and A.~Sa\'a~Hern\'andez\thanksref{addr2}
        \and T.~Sworobowicz\thanksref{addr1}        
        \and C.~T\"urko\u{g}lu\thanksref{addr1}
        \and S.~Williams\thanksref{addr7}
}
\thankstext{e1}{e-mail: mkuzniak@camk.edu.pl}
\thankstext{e2}{e-mail: diego.gonzalez.diaz@usc.es}
\institute{AstroCeNT, Nicolaus Copernicus Astronomical Center of the Polish Academy of Sciences, Rektorska 4, 00-614 Warsaw, Poland\label{addr1}
\and
Instituto Galego de F\'isica de Altas Enerx\'ias, Univ. de Santiago de Compostela, Campus Vida, R\'ua Xos\'e Mar\'ia Su\'rez N\'u\~nez, s/n, Santiago de Compostela, E-15782, Spain\label{addr2}
\and
Institute of Nanostructures, Nanomodelling and Nanofabrication (i3N), Universidade de Aveiro, Aveiro, Portugal\label{addr3}
\and
CERN, Esplanade des Particules 1, Meyrin, Switzerland\label{addr7}
\and
Biological and Chemical Research Centre, Faculty of Chemistry, University of Warsaw, \.Zwirki i Wigury 101, 02-089 Warsaw, Poland\label{addr4}
\and 
Max Planck Institute for Physics, Munich, Germany\label{addr5}
\and
Physics Department, Lancaster University, Lancaster, United Kingdom\label{addr6}
\and
Department of Physics, Technische Universit\"{a}t M\"{u}nchen, 80333 Munich, Germany\label{addr8}
}
\thankstext{currentNikhef}{Currently: Nikhef and the University of Amsterdam, Science Park, 1098XG Amsterdam, Netherlands}
\date{Received: date / Accepted: date}

\maketitle
\begin{abstract}
A new concept for the simultaneous detection of primary and secondary scintillation in time projection chambers is proposed. Its core element is a type of very-thick GEM structure supplied with transparent electrodes and machined from a polyethylene naphthalate plate, a natural wavelength-shifter.
Such a device has good prospects for scalability and, by virtue of its genuine optical properties, it can improve on the light collection efficiency, energy threshold and resolution of conventional micropattern gas detectors. This, together with the intrinsic radiopurity of its constituting elements, offers advantages for noble gas and liquid based time projection chambers, used for dark matter searches and neutrino experiments. Production, optical and electrical characterization, and first measurements performed with the new device are reported.
\keywords{Gaseous electron multipliers
\and Wavelength shifters 
\and Noble gas detectors
}
\PACS{29.40.Mc \and 33.50.Dq \and 95.35.+d \and 14.60.Pq}
\end{abstract}


\section{Introduction}
Dual-phase or gaseous time projection chambers (TPCs) based on noble elements, mainly argon and xenon, are being used for a wide range of fundamental physics experiments, including direct dark matter detection~\cite{ds-50,x1t,lux,ardm}, long baseline neutrino oscillation experiments~\cite{protodune,ariadne}, searches for neutrinoless double beta decay~\cite{Monrabal:2018xlr}, and even applications outside fundamental physics, e.g. in medical or biological imaging~\cite{xemis,SCXM}. 
Energy deposits in such detectors result in primary scintillation (S1) and ionization. By means of a mild electric field, electrons from ionization are collected into a high field region where they induce secondary scintillation (S2) through electroluminescence (EL) or avalanche generation.

A typical configuration, allowing to register both S1 and the ionization component, is based on a cryogenic liquid volume with a gas pocket on top, supplied with wire grid/mesh electrodes defining the high field region, an array of photosensors at the bottom end-cap and, optionally, reflective liner on the side walls. The ionization signal can be registered by collecting the multiplied charge or via S2 scintillation, which for the latter case requires another photosensor array at the top end-cap.

Upcoming and future generation experiments, particularly in the long baseline neutrino oscillation program and for direct dark matter searches, will reach masses of the order of hundred tonnes (ARGO~\cite{ESPPU}, DARWIN~\cite{darwin}) or multi-kilotonne (DUNE~\cite{dune}). While single phase design with `S1-only' readout simplifies the scalability of a detection technology, the simultaneous registration of the ionization channel permits lower energy threshold and accurate reconstruction of the event topologies, giving access to additional physics~\cite{Agnes_2021}. 

Scaling up a dual-phase TPC brings challenges related to the stability and uniformity of charge and light collection, associated with liquid levelling, thermodynamical stability and space-charge effects at the liquid-gas interface (for a recent review of the detection schemes currently considered see e.g.~\cite{LArTPCreview}).
Micropattern gas detectors (MPGDs) such as thick gaseous electron multipliers (THGEMs) were proposed as a scalable charge collection and amplification scheme for large LAr detectors~\cite{BONDAR2007493} and have operated in large scales~\cite{lem,ProtoDUNErun}. Optical readout of THGEMs was also suggested in~\cite{Lightfoot_2009}. Since high light yield and light collection are crucial for achieving the low energy threshold necessary for direct dark matter searches, UV-sensitive (below 400~nm) CsI photocathode coatings were used to make THGEMs not only S2 but also S1-sensitive~\cite{Erdal_2020}, with a a few percent photon detection efficiency (PDE) demonstrated in LXe (for a recent review see~\cite{Buzulutskov}).
Another challenge in collection and detection of S1 and S2 light is its dominant vacuum ultraviolet (VUV) wavelength, 128~nm for LAr and 172~nm (178~nm) for GXe (LXe), strongly absorbed in most materials and in the region where the PDE of available sensors is reduced with respect to visible light. While direct VUV detection yields are satisfactory in the case of LXe, wavelength shifter (WLS) coatings on photosensors and detector walls are necessary to shift the VUV to visible light in case of LAr ~\cite{WlsReview}.

One aspect of particular importance is that, in the EL regime, the optical gain is a linear function of the operating voltage, as \cite{GONZALEZDIAZ2018200}:
\begin{equation}
    m_\gamma = k (V - V_{th}) 
\end{equation}
Here, while $V_{th}$ depends mildly on the amplification gap, $g$, and pressure, $P$, (e.g., $V_{th}=E_{th}^* \cdot g \cdot P$, with $E_{th}^*\simeq0.7$ kV/cm/bar for xenon) it is generally satisfied that $V \gg V_{th}$ and so the EL yield depends largely on the voltage that the structure can sustain, and not on the applied field, unlike for conventional MPGDs. Thus, operation under very thick gaps and pressurized conditions, for which the breakdown voltage of the gas increases, is desired (e.g.,\cite{FREITAS2010205}, \cite{Monrabal:2018xlr}). For experiments where energy resolution is a critical parameter, and presumably for electron-counting too, operation in a pure (avalanche-free) EL regime is non-negotiable \cite{Renner:2019pfe}.

There are currently two types of dedicated structures for EL, both based on radiopure materials, one relying on the transparency of the bulk material (PMMA)~\cite{Gonzalez_Diaz_2020} and another one on its reflective properties (PTFE)~\cite{Ban:2017nnm}. In this work we propose an alternative approach to the FAT-GEM concept (field-assisted transparent gaseous electroluminescence multiplier) introduced in~\cite{Gonzalez_Diaz_2020}, modified in a way which maximizes the S1 light collection and simplifies the wavelength-shifting procedure. 
For increased light collection, the present concept can be easily combined with other proposed schemes such as 3D optical readout~\cite{ariadne}, Liquid Hole Multipliers (LHM)~\cite{Erdal_2020} or THGEMs sensitive to the visible and near infrared component of the gas phase emission~\cite{ds-gem}.

In the following sections the new concept and its possible configurations, fabrication process, a simulation of the main optical properties (yields and point spread function) and first functionality tests of the prototype devices will be introduced.

\section{Concept and possible configurations}
\begin{figure}[ht]
\centering\includegraphics[trim=6cm 3cm 2cm 4cm,clip=true,width=0.8\linewidth]{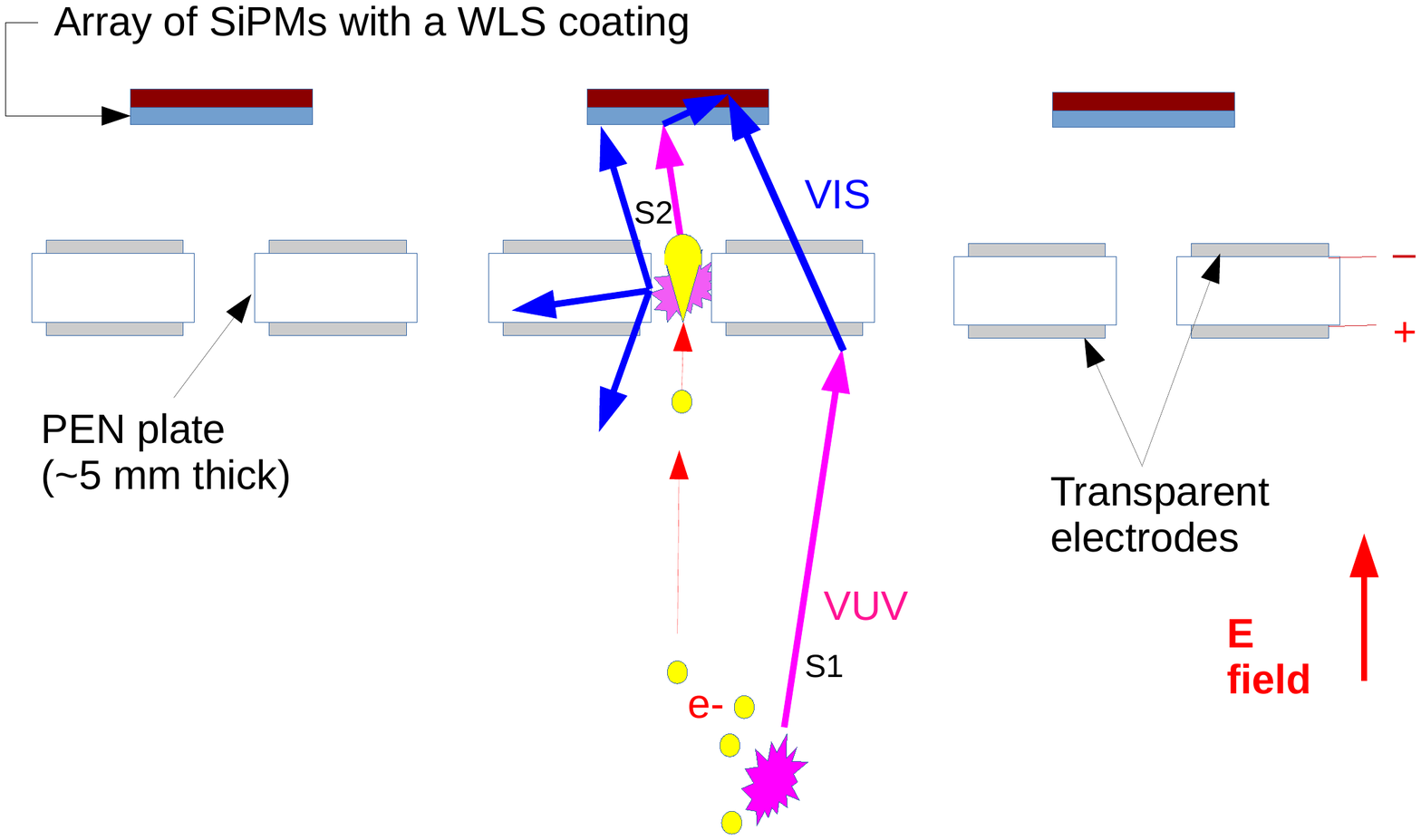}
\includegraphics[trim=6.5cm 6cm 10cm 6cm,clip=true,width=0.53\linewidth]{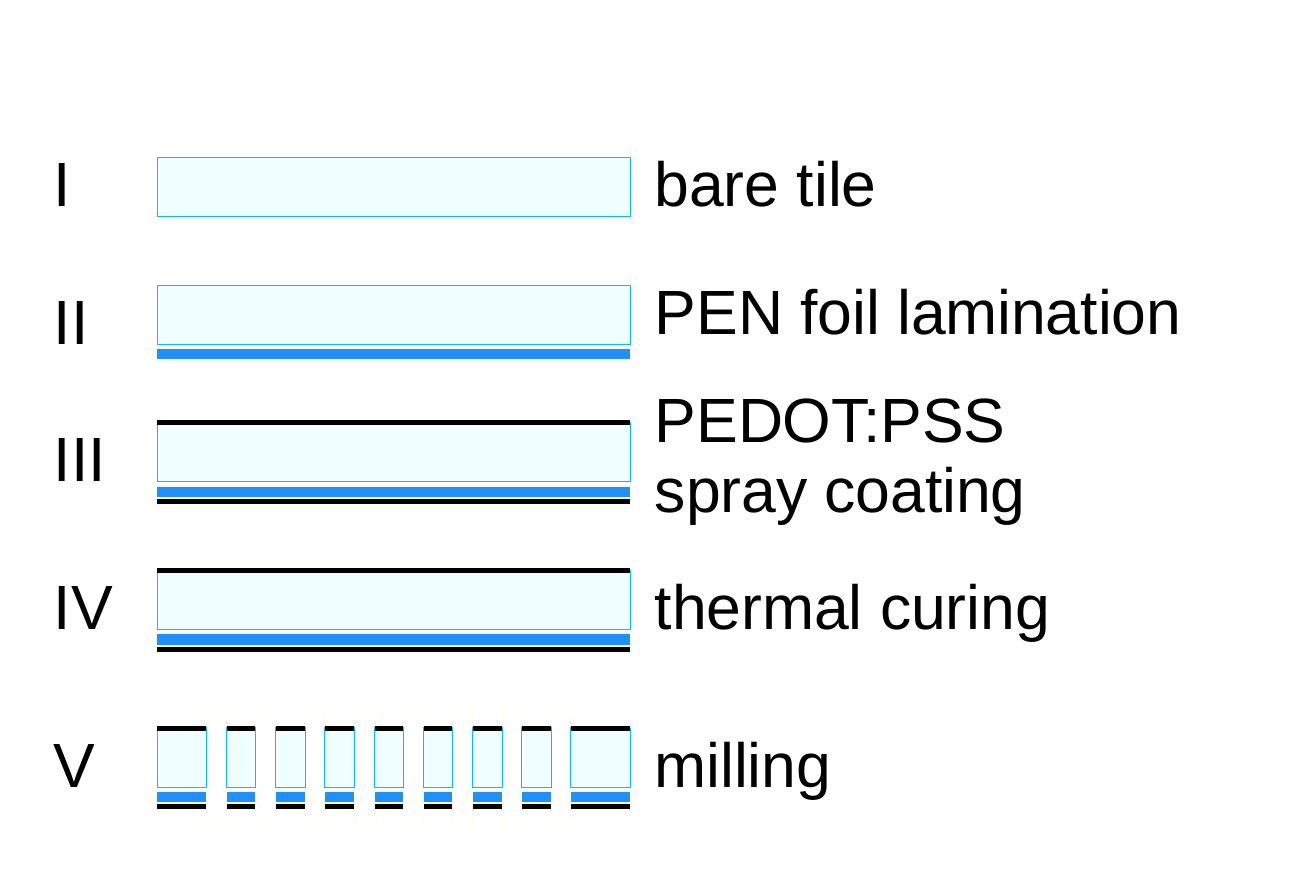}\includegraphics[trim= 26cm 23.7cm 7cm 0cm,clip=true,width=0.47\linewidth]{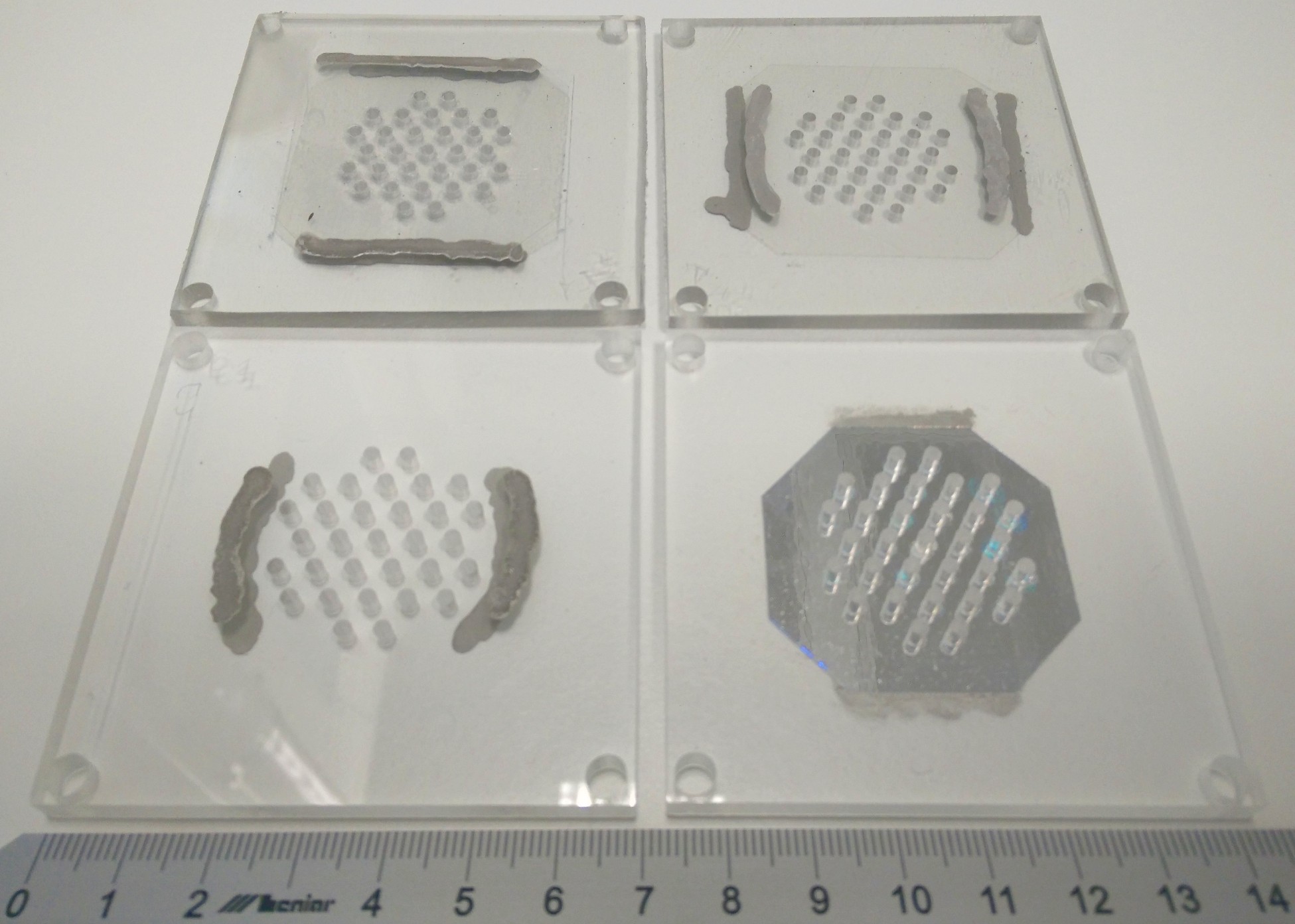}
\caption{(Top) Cross-sectional scheme illustrating the operation of the new WLS-transparent amplification structure, showing photon tracks from S1 and S2 emissions, with VUV in magenta and visible in blue. (Bottom-left) Steps of the fabrication sequence for structure `e'. (Bottom-right) Fabricated structure `e', with deposited silver-based epoxy contacts. }
\label{fig:schematic}
\end{figure}

Main motivations for the amplification structure proposed in Fig.~\ref{fig:schematic}(top) are (1) improving the scalability and stability over traditional meshes/wires currently employed in noble-element TPCs, while at the same time (2) increasing both the EL yield and light collection efficiency over conventional MPGD structures and, finally, (3) wavelength-shifting primary and secondary scintillation to facilitate optical readout.
The first goal is achieved by confining S2 emission to the holes, which makes dual-phase TPCs less sensitive to the gas pocket uniformity and provides at the same time a rugged amplification structure immune to deformation or electrostatic sagging, `tileable' and easy to test in small systems. 
The second goal is achieved by making the amplification structure highly transparent to visible light (i.e., light already wavelength-shifted locally or elsewhere in the detector), as well as making it wavelength-shifting for VUV light stemming from either S1 or S2 scintillation. Collection efficiency for S1 is increased due to the WLS effect at the surface facing the drift region, while for S2 it is increased due to the WLS effect of the holes' walls. This also guarantees the third goal, with the array of photosensors placed after the structure becoming easily sensitive to both S2 and S1 components, which is useful for background mitigation purposes, e.g. for discrimination of pileup of S1-only and S2-only events~\cite{ds-50-full}. Details on the implementations chosen to achieve these goals follow below.

In TPCs, the VUV component of S1 can be collected from the full solid angle by walls covered with an efficient reflector. For xenon scintillation, PTFE provides sufficient reflectivity, while in case of argon a layer of WLS on the reflectors is additionally required (typically TPB-coated ESR or PTFE are used~\cite{WlsReview}). S1 light can therefore reach the amplification structure directly as VUV, or after multiple reflections from the walls, either as VUV or visible photons, if the WLS coating on the walls is present. Transparency to visible light was achieved with PMMA-based structures in~\cite{Gonzalez_Diaz_2020} or polyethylene naphthalate (PEN) base material, supplied with electrodes using either a semi-transparent copper mesh, or a transparent coating of 3,4-ethylene dioxythiophene:poly(sty\-rene sulfonate), PEDOT:PSS~\cite{pedotpssrev}, as in NEXT-White~\cite{Monrabal:2018xlr}. All these materials have low out-gassing characteristics, are vacuum and HV compatible, radiopure, and have been previously used in the context of ultra-low background cryogenic experiments.

Given that PEN has been identified as an efficient wavelength-shifter, and taking advantage of the fact that injection moulding technology for production of up to 5~mm plates has been recently established~\cite{Efremenko_2019}, using it as the bulk material for the amplification structure represents a natural choice. Therefore, several PEN-based tiles have been manufactured to optimize the design and decouple possible light production mechanisms, as shown in Tab.~\ref{tab:config}. In particular, since injection moulded PEN~\cite{Abraham:2021otn} has at the moment a reduced WLS efficiency (WLSE) compared to PEN foils, an additional commercial layer was applied to the surface, aimed at higher S1 sensitivity. PMMA-based structures were prepared as well, for comparison.

\section{Production}
The production of 5~mm-thick injection moulded PEN plates has been developed as part of the R\&D effort for LEGEND~\cite{Efremenko_2020}. The plate used in this work was made starting from a commercial PEN granulate (TN-8065 S) acquired from Tejin-DuPont under the brand name of Teonex~\cite{teonex}.
For configurations `e', `f', `g', additional patches of wavelength-shifting foil (PEN Teonex Q51, 25~$\micro$m-thick, see~\cite{KuzniakPEN}), or specular reflector (ESR from 3M) were laminated on the TPC-side of the tile with an adhesive film, 467MP from 3M (PEN laminates have already been used in LAr~\cite[Sec. 5.16.2.1 therein]{dune}).

Two types of transparent electrodes were tried: (1) a stainless steel or copper mesh with 0.5~mm-pitch and 0.1~mm-wide tracks (71\% transparent), similar to~\cite{Gonzalez_Diaz_2020}, and (2) a thin air-brushed PEDOT:PSS coating, based on Clevios~F~ET (Heraeus). Otherwise, the geometry of the tiles follows the FAT-GEM design~\cite{Gonzalez_Diaz_2020}: 7~cm $\times$ 7~cm$ \times$ 0.5~cm tiles with 2~mm diameter cylindrical holes arranged in a 5~mm pitch hexagonal pattern.
\begin{table}[htb]
\footnotesize
\centering
\begin{tabular}{l|rrcl|cc|cc}
 & \multicolumn{2}{r}{\bf TPC} & {\bf Base} & {\bf PD} &\multicolumn{2}{c|}{\bf S1}& \multicolumn{2}{c}{\bf S2} \\ 
& \multicolumn{2}{r}{\bf {\tiny |}\ \ \ } & {\bf mat.:} & {\bf\ \ {\tiny |}}& \multicolumn{2}{c|}{$\varepsilon_{vuv}$: 1.5\%} & \multicolumn{2}{c}{$\varepsilon_{vuv}$: 10\%}  \\
& \multicolumn{2}{r}{\multirow{2}{*}{\bf side}} & \multirow{2}{*}{\bf\tiny WLSE} & \multirow{2}{*}{\bf side} & {\bf $\varepsilon_{vis}$} & \multirow{2}{*}{\bf $f$} & {\bf $\varepsilon_{vis}$} & \multirow{2}{*}{\bf $f$} \\ 
& \multicolumn{2}{r}{\ \ \ } & {} & {\ \ {\tiny }}&  [\%] &  & [\%] & 
\\
\hline
{\bf a} & \multicolumn{2}{r}{\multirow{4}{*}{\rotatebox[origin=c]{90}{\tiny PEDOT}}} & {\tiny PMMA:0} & \multirow{4}{*}{\rotatebox[origin=c]{90}{\tiny PEDOT}} & 0 & 1 & 0 & 1 \\
{\tiny b} & & & {\tiny PEN:0.09}&  & 1.1(5) & 2.5 & 1.3(5) & 1.2 \\
{\tiny c} & & & {\tiny PEN:1} &  & 12(5) & 17 & 14(5) & 4.9\\
{\tiny d} & & & {\tiny PMMA:1} &   & 32 & 44 & 36 & 8.1 \\
\hline 
{\bf e} & \multirow{3}{*}{\rotatebox[origin=c]{90}{\tiny PEDOT}} & {\tiny Q51} & {\tiny PEN:0.09} & \multirow{3}{*}{\rotatebox[origin=c]{90}{\tiny PEDOT}} & 4(2)& 6.6 & 1.4(6)& 1.2 \\
{\bf f} & & {\tiny Q51} & {\tiny PMMA:0} &   & 2.3 & 4.2 & 0 & 1  \\
{\bf g} & & {\tiny ESR} & {\tiny PMMA:1} &   & 7.7 & 11 & 58 & 13 \\
\hline
{\bf h} & \multicolumn{2}{r}{\multirow{3}{*}{\rotatebox[origin=c]{90}{mesh}}} & {\tiny PEN:0.09} & {\multirow{3}{*}{\rotatebox[origin=c]{90}{mesh}}} & 0.8(3) & 2.1 & 1.0(3) & 1.1 \\
{\tiny i} & \multicolumn{2}{r}{} & {\tiny PMMA:1} &  & 19 & 26 & 25 & 6.0 \\
{\tiny j} &  & & {\tiny Ar/Xe:0} &  & 0 & 60 & 0 & 4.5 \\
\end{tabular}
\caption{Considered and fabricated (bold label) structures, detailing the TPC and photodetector (PD) sides, and base material configurations with the assumed base material WLSE (`g' so far prototyped only with non-WLS PMMA). Light collection efficiencies, $\varepsilon_{vuv}$ (VUV) and $\varepsilon_{vis}$ (visible), obtained from Geant4 simulations (Sect.~\ref{sec:sim}), are given in the right columns (for `j', which is a special case, $\varepsilon^{S1}_{vuv}$=88\% and $\varepsilon^{S2}_{vuv}$=45\%).
For the PEN base material, two visible light absorption lengths are considered, 1~cm and 5~cm, with the resulting central value and systematic uncertainty given for $\varepsilon_{vis}$. The figure of merit, $f$, defined later in text, is also shown (central value only).}
\label{tab:config}
\end{table}

Fabrication steps are shown in Fig.~\ref{fig:schematic}(bottom-left), taking as a reference structure `e'. The production of transparent conductive coatings was performed in an ISO7 cleanroom. Before applying the coatings, the substrates were ultrasonically washed for 1~hour in a solution of Alconox\textsuperscript{\textregistered} in ultrapure water (UPW) at 40$^\circ$C, followed by a ultrasonic wash and rinse in UPW alone and drying. Edges of each tile were then masked with adhesive tape in order to ensure lack of electrical contact with the support structure. After diluting  PEDOT with a solution of isopropanol and ultrapure water (UPW), coatings were deposited using a hand-held commercial airbrush (Paasche Talon TG-3F), supplied with a fan head permitting to uniformly deposit up to a 3-inch wide beam. The target thickness of the coatings, approximately 50~nm, was achieved by adjusting the amount of PEDOT in the solution relative to the area of the substrate. Deposition was performed in three passes for each substrate side. Coatings were then cured in an oven at 85$^\circ$C for 30~minutes. The behaviour of the coating down to LN$_2$ temperature, for two different target values of the surface resistivity, was measured with a purposely-designed setup at IGFAE, showing an increase in resistivity by a factor of approx.~50 (3) between room temperature and 77~K for the 20~M$\mathrm{\Omega}$/sq. (5~k$\mathrm{\Omega}/$sq.) coating, see Fig.~\ref{fig:spectra}(top). While this aspect is of importance for detector stability and spark immunity, its optimization was not part of the present study. Coatings used were in the range of M$\mathrm{\Omega}$/sq.

For the case of transparent electrodes based on PEDOT (structures `a', `e', `f'), the coating was applied directly to the PEN and PMMA tiles (with additional Q51 and ESR films depending on the case), and the holes milled out. Finally, electrical leads were attached to the sides of the transparent electrodes with silver-based epoxy (Technicqll R-082), see Fig.~\ref{fig:schematic}(bottom-right). Apart from the production process, the PEN and Teonex materials and the final tiles were stored in the dark.

For the case of semi-transparent electrodes based on meshes (structure `h'), two different methods were attempted: thermal bonding of copper-substrate followed by photolithographic etching of the mesh (at RD51 workshops, as in \cite{Gonzalez_Diaz_2020}) and 
contact by pressure with stainless steel woven meshes.

\section{Expected light collection performance}\label{sec:sim}
A Geant4~\cite{geant4} simulation model was used to quantify light collection. Optical properties were taken from the literature, including: PEN and PMMA attenuation lengths of 1--5~cm~\cite{Efremenko_2020} and $\sim$3~m~\cite{deap}, respectively, PEN emission spectrum~\cite{KuzniakPEN}, refractive indices of PEN and PMMA, and ESR reflectivity of 0.98.
The literature results for PEN WLSE, are typically relative to TPB, which itself has a debated absolute yield~\cite{WlsReview}. For the sake of this discussion we treat the published PEN WLS efficiencies as if they were absolute: 9.1\% for the PEN tile and 34\% for Teonex Q51 film, as recently measured for a similar grade in a LAr environment~\cite{Abraham:2021otn}. Assumed WLSE are constant in the entire VUV range, which again is debated~\cite{Graybill:2020xhu,Benson} and makes the results nominally valid for both Ar and Xe and easy to scale to a more realistic WLSE in presence of new data. 

\begin{figure}[ht]
\centering
\includegraphics[trim=1cm 0cm 0cm 0cm,clip=true,width=\linewidth]{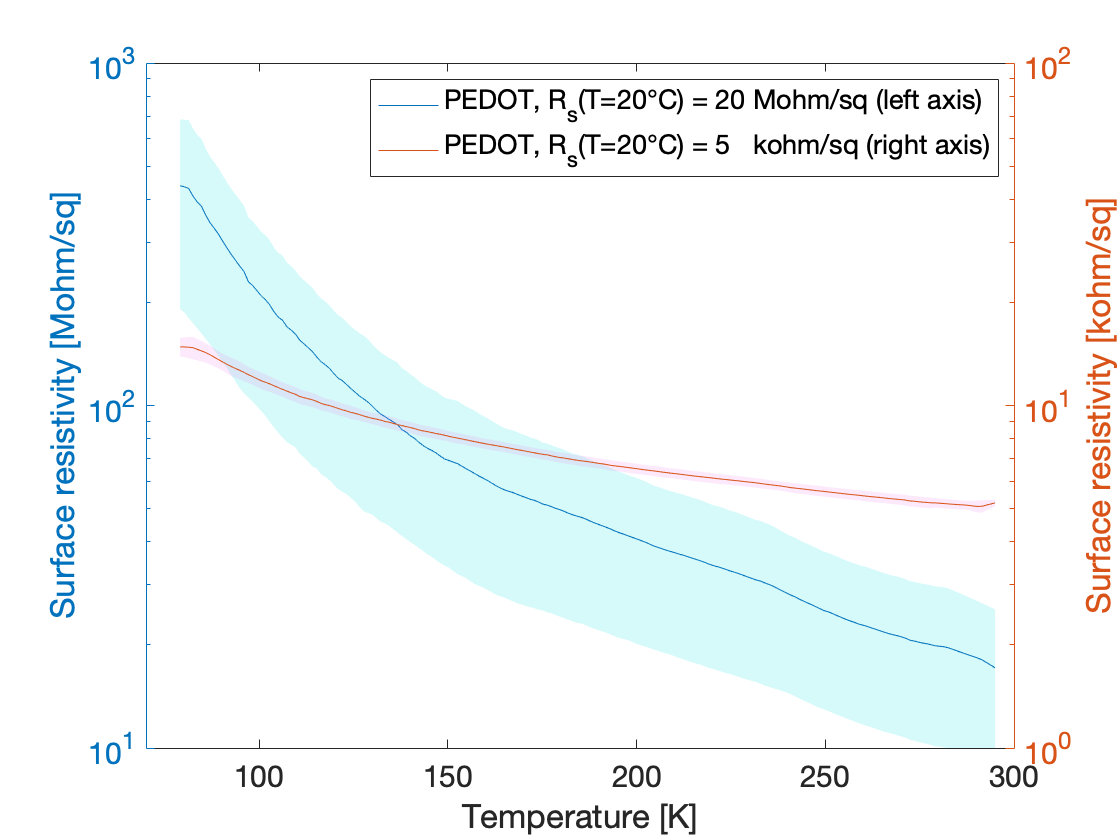}
\includegraphics[width=\linewidth]{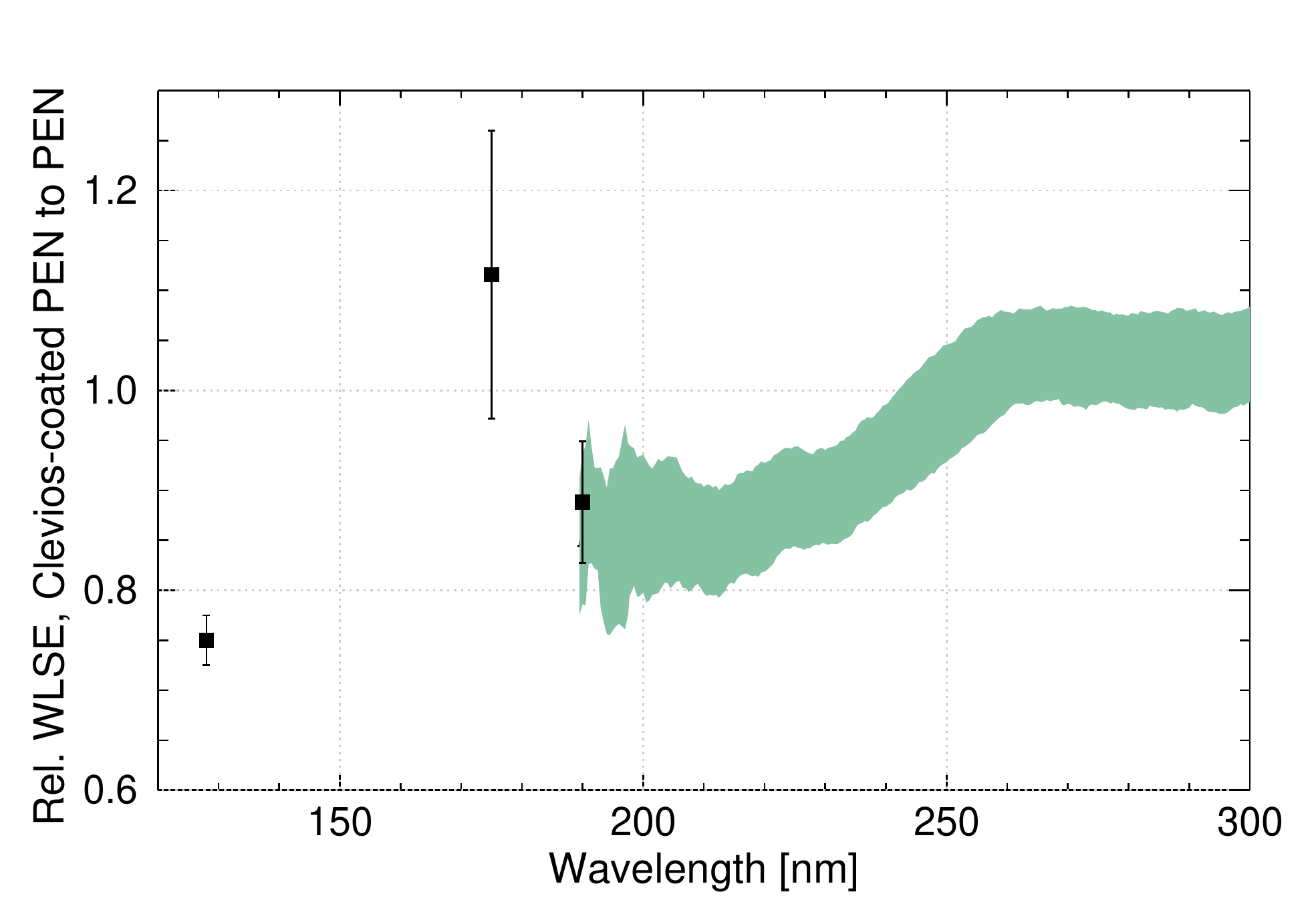}
\caption{Characterization of the PEDOT coating. (Top) Measurements of the sheet resistivity vs. temperature for two extreme cases, obtained from the average on different spots. Surface resistivities on the range of 10-1000~M$\mathrm{\Omega}$/sq are interesting because they offer spark quenching, however they are thinner and less uniform (illustrated by the larger error band). (Bottom) Ratio of WLSE measurements of PEN foils with and without PEDOT electrodes. Performed using a Shimadzu UV-3600 spectrophotometer equipped with an integrating sphere in a diffuse reflection configuration (green band) and with a fixed-detection-angle vacuum monochromator setup (black points).}
\label{fig:spectra}
\end{figure}

The least known parameter is the transparency of PEDOT coatings. On the one hand, technical specifications state up to 99\% for visible light~\cite{clevios}, which is used in the simulation. On the other hand, VUV-transparency is not known. In order to assess this effect, an integrating sphere (LISR-3100) measurement in a diffuse reflectivity mode was performed with a Shimadzu UV-3600 spectrophotometer down to 190~nm, and with a fixed-detection-angle vacuum monochromator (Acton VM-502) setup down to 128~nm. Following the method described in~\cite{KuzniakPEN}, we evaluated the apparent WLSE reduction caused by VUV losses in the conductive coating, see Fig.~\ref{fig:spectra}(bottom). The more conservative 128~nm value of 75\% was chosen, also due to the large uncertainty on the 175~nm data point (caused by a low intensity of the light source at that wavelength).

For reference, we consider some limiting cases, assuming 100\% WLSE for PEN or PMMA (`c', `d', `g', `i'). Such a situation could be approached in practice by optimizing the base material properties, or application of thin evaporated, painted or spin-coated~\cite{Graybill:2020xhu} TPB layers covering the plate surfaces. 
Additionally, configuration `j' in Tab.~\ref{tab:config} approximates the optical configuration of a dual-phase TPC, with a gas pocket replacing the tile and placed between 95\% and 98\% transparent meshes - inspired by DarkSide-50~\cite{ds-50}, but adapted for the considered geometry, for comparison.

The light collection efficiency of the structures was simulated for both S1 and S2 components. For S1, photons were generated isotropically into the cone containing the hole pattern from a point fixed centrally 15~mm from the tile, into the TPC side (1.80~sr solid angle). S2 photons were generated isotropically from the central FAT-GEM hole, with vertex positions following a 2D Gaussian distribution centered at the hole axis, which follows from simulations of the electron transport and generation of excited states in the gas performed with the Garfield++ toolkit~\cite{garfield}.
Collection efficiencies were calculated separately for VUV and visible photons by dividing the number of photons reaching a plane located 3~mm away, on the photodetector (PD) side of the FAT-GEM tile, by the number of initially generated VUV photons. With such definition, photons leaving the tile in the opposite direction are lost; although in a real detector they would still have a chance to be detected on the opposite side of the TPC (or reflected back), improving the overall light yield and energy resolution, they would contribute little information to $x$-$y$ position reconstruction. VUV collection efficiencies are driven by the solid angles and approximately the same for all considered configurations (except for `j'): $\varepsilon^{S1}_{vuv}$=1.5\% and $\varepsilon^{S2}_{vuv}$=10\%.  $\varepsilon^{S1}_{vis}$ and $\varepsilon^{S2}_{vis}$ strongly depend on the optical properties of the model and are summarized in Tab.~\ref{tab:config}.

It is observed that $\varepsilon^{S1}_{vis}$ scales with the PEDOT transparency and the WLSE of the face material, while $\varepsilon^{S2}_{vis}$ scales with the WLSE of the bulk material. The efficiency increases by a factor of approx.~2 with the PEN absorption length increased from 1 to 5~cm, and by another factor of 2 when switching to PMMA, which has negligible absorption losses. We have verified that the conclusions are robust against changing the S2 vertex distributions inside the FAT-GEM hole to uniform (which results in a 9\% reduction of $\varepsilon^{S2}_{vuv}$ and 2\% increase of $\varepsilon^{S2}_{vis}$), and for VUV Ar and Xe scintillation.

Compared to $\varepsilon_{vuv}$, even the simplest WLS configuration, `b', additionally brings a matching amount of visible light for S1, yet approximately eight times lower for S2. With enhanced WLS properties, the number of visible photons can exceed the VUV one by up to a factor of 19 for S1 (structure `d') and up to a factor 6 for S2 (structure `g', which makes use of ESR reflector to maximize S2 collection).

For a more practical comparison one should also consider the typical SiPM photodetection efficiencies (PDEs) of 50\% for blue visible light, significantly higher than 20\% for VUV~\cite{Gola_2019}. We assume, for simplicity, a blue-sensitive sensor coated with a perfect WLS layer, effectively reducing $\varepsilon_{vuv}$ by a factor of two due to isotropic re-emission of wavelength-shifted light. Consequently, a figure-of-merit parameter, representing the enhancement in optical output relative to a non-WLS configuration `a', is defined as:
\begin{equation}
\centering
f\equiv\frac{\varepsilon_{vuv}/2 +\varepsilon_{vis} }{\varepsilon^{(a)}_{vuv}/2},
\end{equation}
Based on this metric, and relying on the current optical properties of PEN plates, it is possible to anticipate an increase of the photoelectron yield by a factor of approximately 2 for S1 and 20\% for S2, with respect to conventional FAT-GEMs (structure `b'). In the limiting cases of WLSE approaching one, the enhancement factor could reach up to 44 for S1 (only 35\% shy of the value achievable for a meshes) and 8 for S2, as in structure `d' in Tab.~\ref{tab:config} -- almost twice the value achievable with meshes (represented by configuration `j'). Such a significant improvement has a purely geometrical origin: in a FAT-GEM configuration, S2 VUV light emitted into nearly full solid angle is isotropically wavelength-shifted, with up to a half reaching PDs, while in the case of meshes only half of the S2 VUV emission can reach the WLS-coated PDs, with only a quarter left after conversion. If a reflector foil is added to the TPC-side of the FAT-GEM, as in structure `g', nearly all wavelength shifted light can be collected (at the expense of loosing S1 light), resulting in the optical gain tripled relative to meshes (as long as the same WLSE can be achieved in both cases).
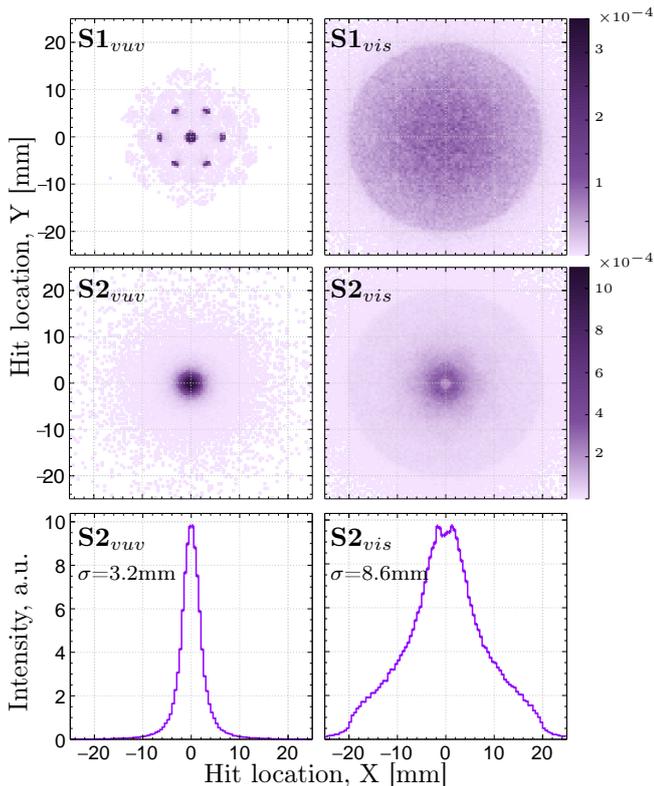
\begin{figure}[ht]
\centering
 \begin{overpic}[width=0.9\linewidth]{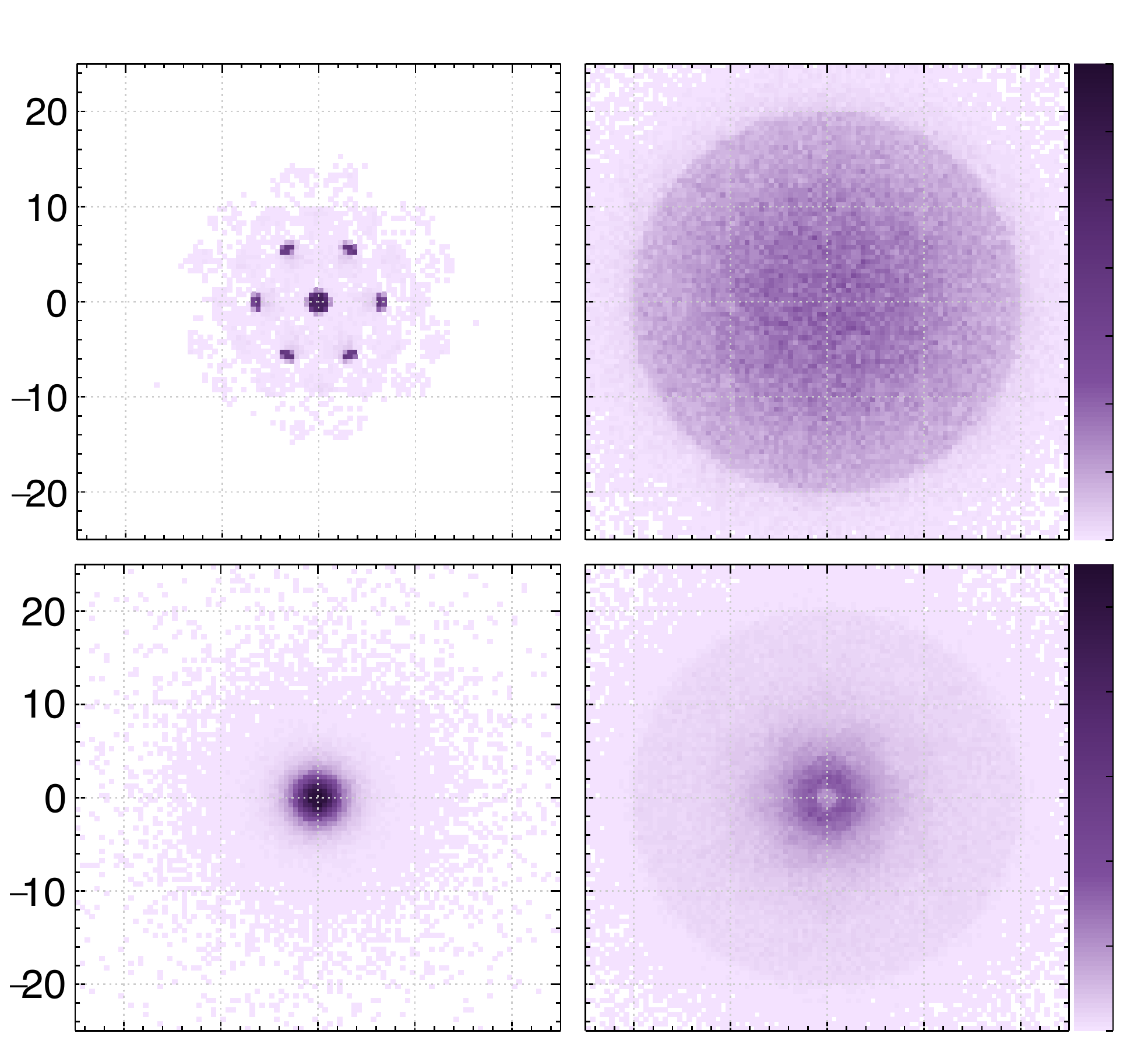}
    \put(98,38){\tiny 10}
    \put(98,30.5){\tiny 8}
    \put(98,23){\tiny 6}
    \put(98,16){\tiny 4}
    \put(98,9){\tiny 2}
     \put(98.5,42.5){\tiny $\times10^{-4}$}   
    \put(98,56.5){\tiny 1}
    \put(98,68){\tiny 2}
    \put(98,80){\tiny 3}
    \put(98.5,85.5){\tiny $\times10^{-4}$}
  \put(-4,23){\rotatebox{90}{Hit location, Y [mm]}}
   \put(8,81){{\bf S1$_{vuv}$}}
    \put(52,81){{\bf S1$_{vis}$}}
      \put(8,37){{\bf S2$_{vuv}$}}
        \put(52,37){{\bf S2$_{vis}$}}
 \end{overpic}
 \begin{overpic}[width=0.9\linewidth]{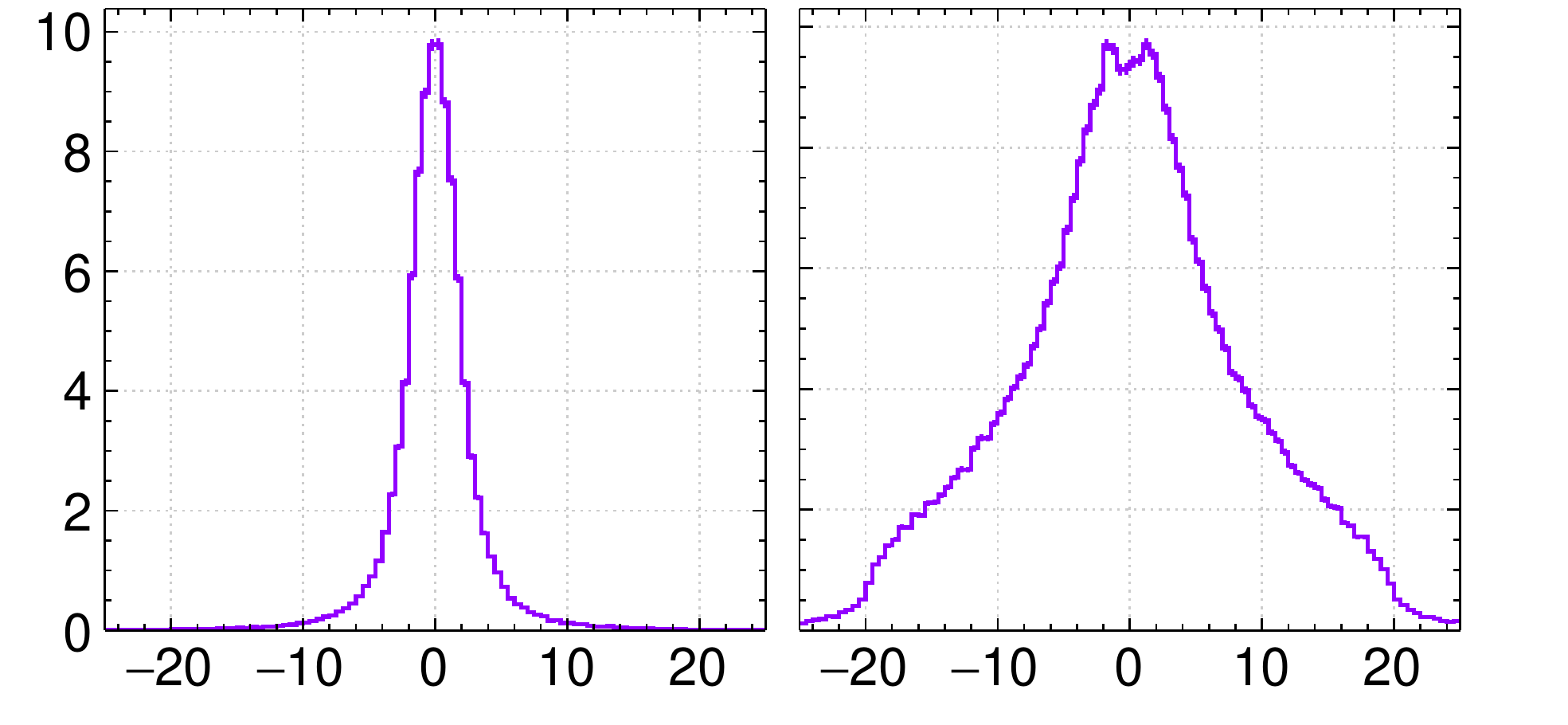}
 \put(30,-1.5){Hit location, X [mm]}
   \put(-4,11){\rotatebox{90}{Intensity, a.u.}}
       \put(8,40){\bf S2$_{vuv}$}
       \put(8,34){\small $\sigma$=3.2mm}
        \put(52,40){\bf S2$_{vis}$}
        \put(52,34){\small $\sigma$=8.6mm}
 \end{overpic}
\caption{Photon hit locations at a photodetection plane placed 3~mm after the WLS FAT-GEM. Shown separately: $x$-$y$ maps for VUV (left column) and visible (right column), photons from S1 (top) and S2 (middle), and $x$-axis projections of S2 hits (bottom). The standard deviation, $\sigma$, is given for S2 projections. The histograms are divided by the initial 10$^6$ of generated VUV photons and have 0.5~mm-wide bins. 
Clusters of VUV hits from photons transmitted directly through GEM holes are visible on top of a broader distribution caused by diffuse reflections.}
\label{fig:xy}
\end{figure}

Finally, distributions of photon hits in an $x$-$y$-plane placed 3~mm after the tile are shown in Fig.~\ref{fig:xy}(top and middle), together with the point spread function (PSF) for S2 (bottom). The visible photon PSF is 2-3 times wider than for VUV, yet recent results show that optical PSFs even wider than the one in Fig.~\ref{fig:xy}(bottom-right) are compatible with track reconstruction with 1-2~mm accuracy~\cite{Simon:2021yab}.

\section{High voltage tests and validation in xenon gas}
\begin{figure}[ht]
\centering
\includegraphics[width=0.50\linewidth]{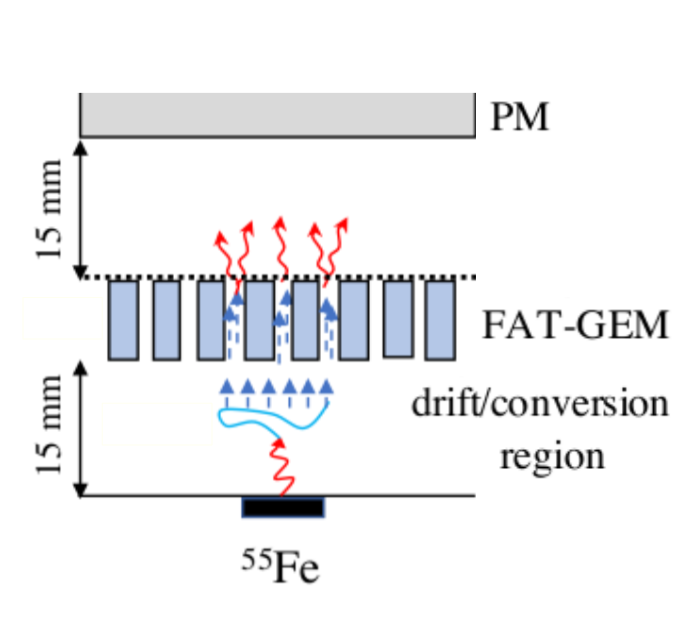}
\includegraphics[trim=5.2cm 9.3cm 5.2cm 9.5cm,clip=true,width=0.48\linewidth]{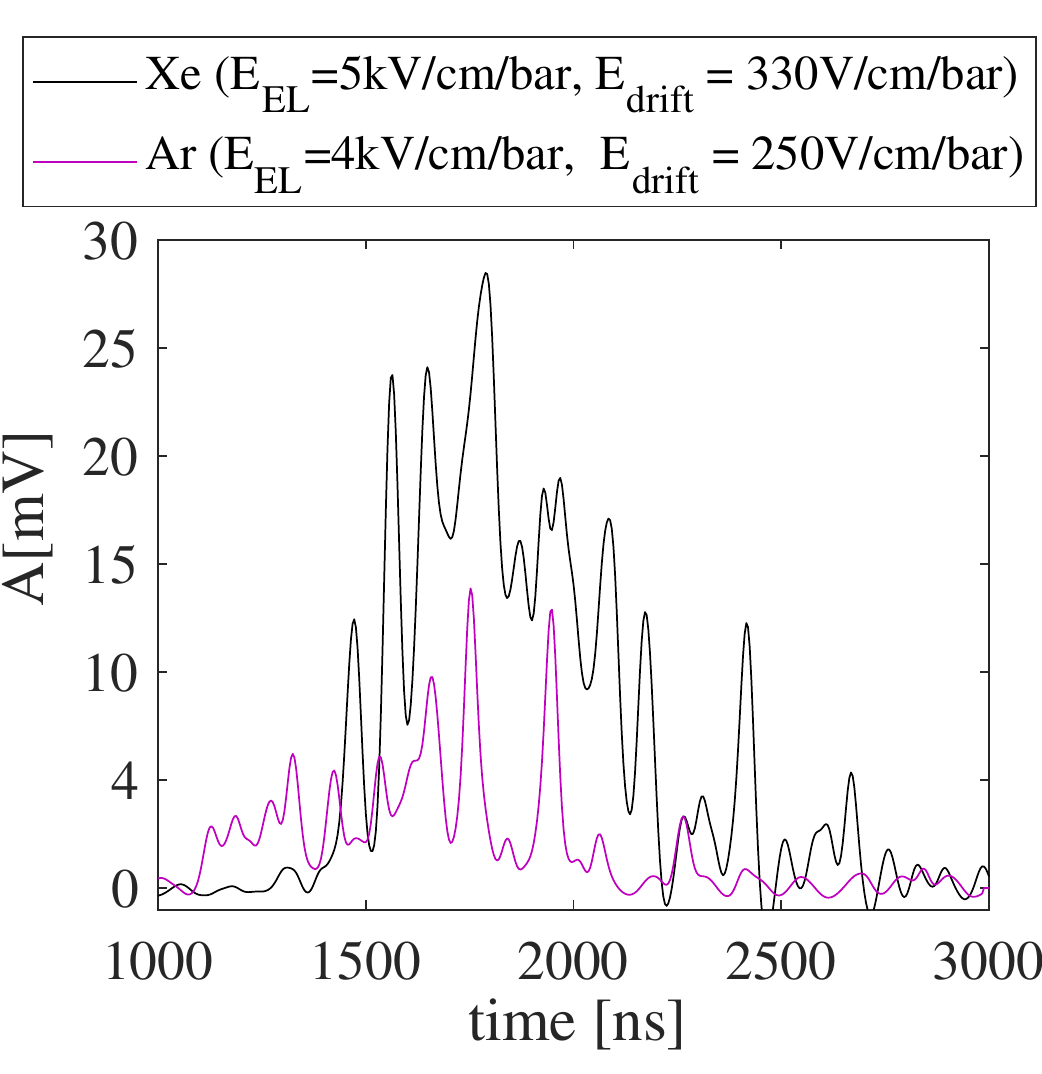}

\includegraphics[trim=5.2cm 9.3cm 5.5cm 9.5cm,clip=true,width=0.49\linewidth]{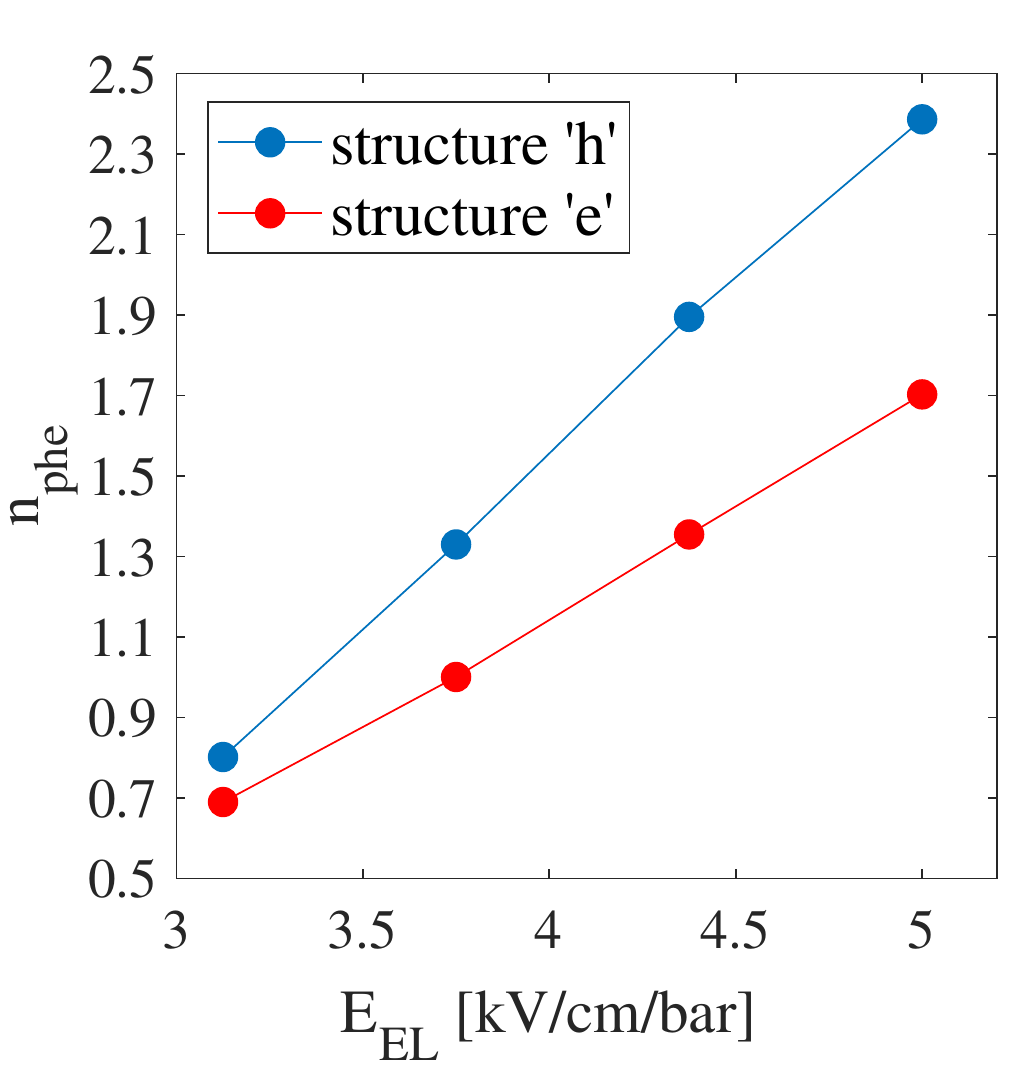}
\includegraphics[trim=5.2cm 9.3cm 5.5cm 9.5cm,clip=true,width=0.49\linewidth]{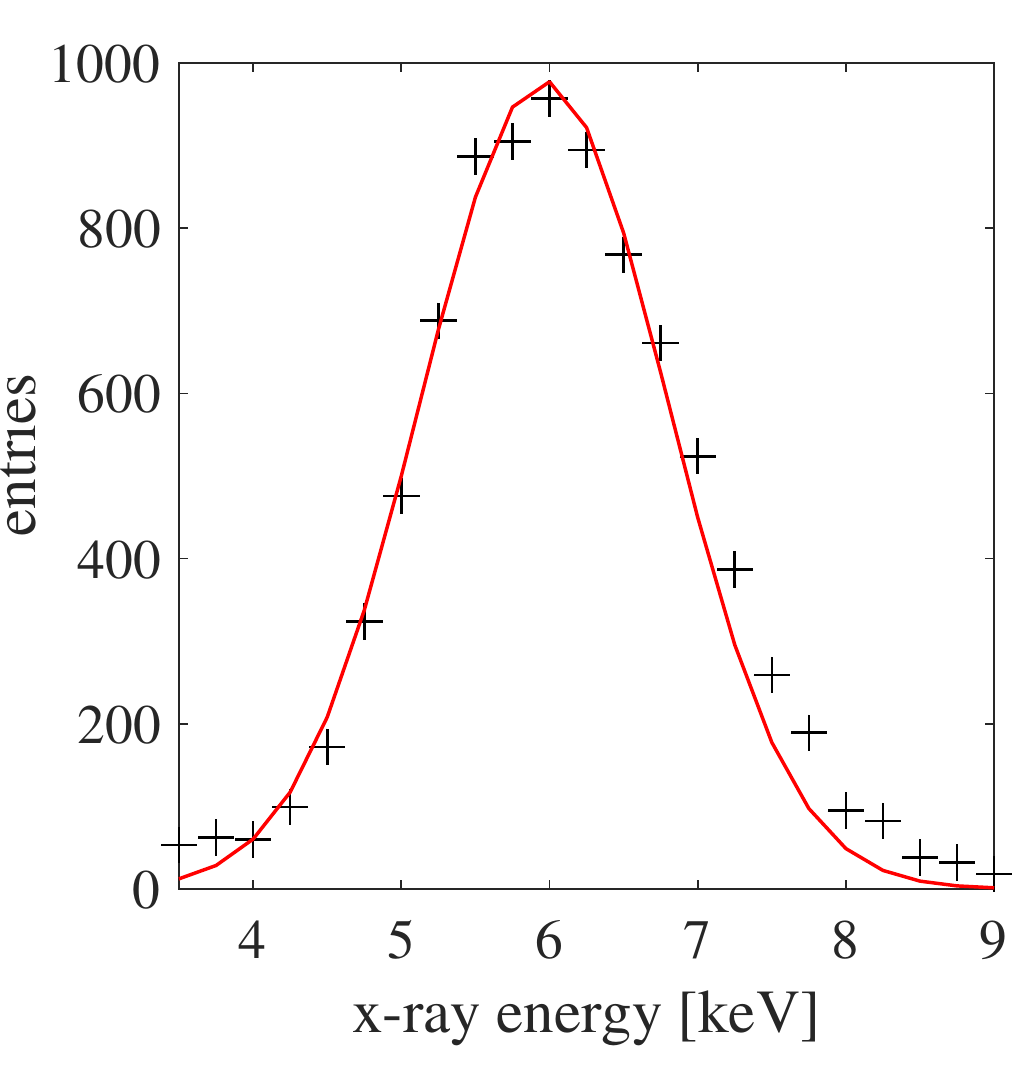}
\caption{(Top-left) Experimental setup used at IGFAE for the characterization of the WLS FAT-GEMs~\cite{Gonzalez_Diaz_2020}. (Top-right) X-ray waveforms for xenon (2~bar) and argon (4~bar) after cuts. (Bottom-left) Light yield in photoelectrons per electron for xenon at 2~bar. (Bottom-right) Reconstructed energy spectrum at a reduced field in the EL region of 5~kV/cm/bar, for structure `e'.}
\label{fig:elresults}
\end{figure}

High voltage stability was first tested in room air, with no adverse effects observed up to 6~kV drop across the structure. In xenon at 2~bar, conditions used for characterization, 4~kV voltage drop could be applied reliably. The setup employed was the same one used in Ref.~\cite{Gonzalez_Diaz_2020}, see Fig.~\ref{fig:elresults}(top-left). X-ray events from an $^{55}$Fe source were produced at the cathode, with a characteristic mean free path around 1.5~mm in xenon at 2~bar (5~mm in argon at 4~bar). Since most events come from the cathode, the shape of the waveforms is diffusion dominated, see Fig.~\ref{fig:elresults}(top-right). The PMT used (PM7378) is blind to argon scintillation, so the relatively high yields observed in this latter case hint at WLS taking place on the PEN plate. The behaviour of the optical gain (in photoelectrons/electron) is shown in Fig.~\ref{fig:elresults}(bottom-left) for structures `h' and `e', and the energy spectrum corresponding to the highest EL field used
for structure `e' is shown in Fig.~\ref{fig:elresults}(bottom-right), giving 32\% (FWHM). These values are comparable to the ones obtained previously for acrylic structures in this setup, for this gas, field and pressure conditions. They cannot be interpreted directly in terms of those achievable in a final experiment, however, since the purity levels, sensor type and geometrical coverage are setup dependent. They will be discussed in detail in a forthcoming publication. 

This is to our knowledge the first instance of a micropattern gaseous detector directly bulked on a WLS material. While further optimization is needed, including the WLSE itself, this work shows that the concept is sound, not being limited by surface discharges, outgassing or charge collection/loss at the channels.

\section{Summary}
Transparent/wavelength-shifting amplification structures based on PEN and PEDOT:PSS (Clevios~F~ET) for the simultaneous light and charge detection in noble gas and dual-phase detectors can bring substantial improvements over conventional wires/meshes as well as other MPGD readouts. As shown by the simulation results based either on the literature or directly measured optical parameters, 20\% more S2 photoelectrons and several times more S1 compared to PMMA-based structures can be anticipated, if resorting to injection moulded PEN plates developed as part of the R\&D program of the LEGEND experiment. With more efficient WLS, and a base material with negligible losses, an order of magnitude enhancement for S2 (and a factor 44 for S1) could be obtained. It is in principle possible to achieve such a performance with commercial WLS plates or optimized configurations of PMMA base plates coated with PEN or TPB, that allow WLSE's in the range 0.5-1, depending on the source. Comparison with a TPC supplied with wire mesh electrodes, on the other hand, show that such optimized structures could achieve a similar S1-light collection yield (up to 75\%) and up to 2-3 times higher for S2, due to a much more favourable geometry for collection and wavelength shifting of VUV electroluminescence.

Several transparent/WLS amplification structures (with 3.5~cm diameter of active region) were made, using mesh or PEDOT electrodes, showing for the latter: good scalability potential, adequate sheet resistivity for temperatures between 87~K and room temperature, and high transparency for VUV light. High voltage tests and a proof-of-principle run in 2~bar gaseous Xe TPC were performed, permitting to register first light from the structures, hints of wavelength shifting, and demonstrating adequate performance of the PEDOT coatings.

Future plans include the detailed empirical characterization of S1 and S2 collection efficiencies, further HV optimization of the transparent electrodes, implementation of the TPB-doped polymeric coatings and longer term and larger scale performance and stability tests. Based on the projections presented in this work, and the promising first tests, future optimized structures could offer sufficient S1 yield, better S2 yield and, above all, a higher scalability and uniformity of response, needed by the upcoming generations of large noble-element based detectors. 

\section*{Acknowledgements}
This project was funded by the National Science Centre, Poland (Grant No.~2019/03/X/ST2/01560). The NCAC PAS team acknowledges support from the International Research Agen\-da Programme AstroCeNT (MAB/2018/7)
funded by the Foundation for Polish Science from the European Regional Development Fund (ERDF). DGD acknowledges Ramon y Cajal program (Spain) under contract number RYC-2015-18820. AstroCeNT and TUM cooperation is supported from the EU’s Horizon 2020 research and innovation programme under grant agreement No 952480 (DarkWave project). The prototype construction was carried out with the use of CEZAMAT (Warsaw) cleanroom infrastructures financed by the ERDF; we thank Maciej Trzaskowski of CEZAMAT for support.
We are grateful to Prof. Magdalena Skompska for access to the spectrophotometer, which was purchased by CNBCh (University of Warsaw) from the project co-financed by EU from the ERDF.
We are grateful to Robert Hetma\'nski and Pawe\l\ Rozi\'nski of Deadline Pracownia S.\,C. for machining the FAT-GEM plates.
We acknowledge fruitful discussions with Piotr Biel\'owka, Mark Boulay, Jeffrey Mason, Francesco Pietropaolo, Edgar Sanchez, Masayuki Wada and Hanguo Wang. Discussions with the PEN consortium members, including Ryan Dorrill, Michael Febbraro, Brennan Hackett, Bryce Littlejohn and Bela Majorovits, particularly on details of PEN machining, are gratefully acknowledged. We thank the RD51 collaboration for supporting this project at an early stage, and providing resources to fabricate the PEN-based tiles with micropatterned copper mesh.

\bibliography{pengem}
\bibliographystyle{spphys}
\end{document}